\documentclass[12pt]{article}
\usepackage{amsmath,amssymb,theorem,cite,epsfig,url,psfrag,eepic,mathtools,amsmath}
\mathtoolsset{showonlyrefs}
\advance \topmargin by -\headheight
\advance \topmargin by -\headsep     
\evensidemargin \oddsidemargin
\def\Maketitle{{\def\newpage{}\maketitle}}
\begin{document}
\title{\textbf{Integrable $\mathfrak{gl}(n|n)$ Toda field theory\\ and its sigma-model dual}\vspace*{.3cm}}
\date{}
\author{A. V. Litvinov\\[\medskipamount]
\parbox[t]{0.85\textwidth}{\normalsize\it\centerline{Landau Institute for Theoretical Physics, 142432 Chernogolovka, Russia}}}
\Maketitle
\begin{abstract}
In these notes we study the duality between sigma-models and Toda QFT's.  We claim that $\mathfrak{gl}(n|n)$ affine Toda field theory behaves in the strong coupling limit as $\eta-$deformed $\mathbb{CP}(n-1)$ sigma-model plus a free field.
\end{abstract}
Duality is an interesting concept of modern theoretical physics \cite{Polchinski:2014mva}. Especially  the weak/strong coupling one. It replaces strongly interacting regime of one theory with perturbative regime of the other, and vise versa. In these notes we study the duality between  integrable Toda QFT's based on supergroups and $\eta$-deformed sigma-models along the lines suggested in \cite{Fateev:2018yos,Litvinov:2018bou}. We consider  $\mathfrak{gl}(n|n)$ affine Toda field theory \cite{Litvinov:2016mgi,2015arXiv151208779B}
\begin{equation}\label{action}
   \mathcal{A}_{n}=\int\left(\frac{1}{8\pi}\left(\partial_{a}\Phi\cdot\partial_{a}\Phi\right)+\frac{1}{8\pi}\left(\partial_{a}\phi\cdot\partial_{a}\phi\right)+
   \Lambda\sum_{k=1}^{n}\left(e^{b\Phi_{k}-i\beta\phi_{k}}+e^{i\beta\phi_{k}-b\Phi_{k+1}}\right)\right)\,d^{2}\xi,\quad\beta=\sqrt{1+b^{2}},
\end{equation}
where $\phi=(\phi_{1},\dots,\phi_{n})$, $\Phi=(\Phi_{1},\dots,\Phi_{n})$ are two $n-$component bosonic fields and $\Phi_{n+1}\overset{\text{def}}{=}\Phi_{1}$. The parameter $b$ plays the role of the coupling constant of the theory. Each exponent in \eqref{action} has fermionic scaling dimensions $\Delta=\bar{\Delta}=\frac{1}{2}$ and hence $\Lambda$ has a dimension of mass. The theory \eqref{action} contains  the $U(1)$ part $\chi=\beta\sum\Phi_{k}-ib\sum\phi_{k}$, which is not interacting. In order to make it manifest, we perform the following transformation of the fields (here $Z=\sum_{k=1}^{n}(\Phi_{k}+i\phi_{k})$)
\begin{equation}\label{Z-Z}
   Z\rightarrow (\beta+b)Z,\quad\bar{Z}\rightarrow(\beta-b)\bar{Z}.
\end{equation}
After this transformation the action will have the form
\begin{equation}\label{action-modified}
   \mathcal{A}_{n}'=\int\left(\frac{1}{8\pi}\left(\partial_{a}\Phi,\partial_{a}\Phi\right)+\frac{1}{8\pi}\left(\partial_{a}\phi,\partial_{a}\phi\right)+
   \Lambda\sum_{k=1}^{n}\left(e^{b(h_{k},\Phi)-i\beta(\mathfrak{h}_{k},\phi)}+e^{i\beta(\mathfrak{h}_{k},\phi)-b(h_{k+1},\Phi)}\right)\right)\,d^{2}z,
\end{equation}
where $h_{k}=e_{k}-\frac{1}{n}(e_{1}+\dots+e_{n})$ and $\mathfrak{h}_{k}=e_{k}-\frac{\beta-1}{n\beta}(e_{1}+\dots+e_{n})$ and $e_{k}$ is the orthonormal basis in $\mathbb{R}^{n}$. From \eqref{action-modified} it is clear that the ``center of mass'' field $\sum_{k}\Phi_{k}$ decouples. 

In order to define QFT \eqref{action} correctly, one has to specify the domain of the coupling constant $b$. One distinguishes between the weak $b\rightarrow0$ and the strong-coupling $b\rightarrow\infty$ regimes. At $b\rightarrow0$ one has to add an additional UV regularization term
\begin{equation}
  \frac{\pi\Lambda^{2}a^{2}}{b^{2}}\int\sum_{k=1}^{n}e^{b(\Phi_{k}-\Phi_{k+1})},
\end{equation}
where $a$ is the UV cut-off. The model defined by the action \eqref{action} possesses perturbative integrability in the mass parameter $\Lambda$.  In the limit $\Lambda\rightarrow0$ one can construct an infinite tower of local integrals of motion of all spins $(\mathbf{I}_{s}^{(0)},\bar{\mathbf{I}}_{s}^{(0)})$, $s=1,2,\dots$
\begin{equation}
   \mathbf{I}^{(0)}_{s}=\oint\left(\sum_{k=1}^{n}\left(b^{s-2}(\partial\Phi_{k})^{s}+(i\beta)^{s-2}(\partial\phi_{k})^{s}\right)+\dots\right)d\xi,
\end{equation}
and similar expression for $\bar{\mathbf{I}}^{(0)}_{s}$. One can write explicit formulae for lowest integrals, but they are quite cumbersome. We note that there exists an analog of quantum Miura transformation for this theory, which was found in \cite{Prochazka:2018tlo}. Using it, one can in principle find a convenient expression for IM's.  For us it is important that the system of IM's $\mathbf{I}^{(0)}_{s}$  can be defined as a commutant of screening charges
\begin{equation}\label{Scr-charges}
   \mathcal{S}_{k}=\oint e^{b\Phi_{k}-i\beta\phi_{k}}d\xi,\quad \tilde{\mathcal{S}}_{k}=\oint e^{i\beta\phi_{k}-b\Phi_{k+1}}d\xi,\quad k=1,\dots,n,
\end{equation}
which correspond to the exponential terms in the action. We stress that $\mathbf{I}^{(0)}_{s}$ are defined in the theory of  free massless bosonic fields.  They are images of total IM's  at the limit $\Lambda\rightarrow0$. Going beyond the leading order in $\Lambda$ is a quite complicated task, but it is believed that the full theory is integrable.

Another test of integrability comes from the perturbative analysis at $b\rightarrow0$ while keeping $\Lambda$ fixed. It is convenient to use Coleman-Mandelstam duality and replace bosonic fields $\phi_{k}$ by Dirac fermions $\psi_{k}$. This replacement leads to the theory of $n$ Dirac fermions and $n$ bosonic fields $\Phi_{k}$ with hidden $\mathfrak{gl}(n)$ symmetry. It can be shown perturbatively in the parameter $b$ that the scattering of fundamental particles in this theory shares the properties of factorized scattering \cite{Fateev-cpn}. In particular, one finds an absence of particle production -- a remarkable property of integrable QFT's. The exact $S-$matrix for this theory has been conjectured recently in \cite{Fateev-cpn}.

In the strong coupling regime $b\rightarrow\infty$ the action \eqref{action} is useless. However, one can use the following observation. Each pair of fermionic screening charges \eqref{Scr-charges} defines the conformal algebra of the coset CFT $SU(2)_{\kappa}/U(1)$ with $\kappa=-2-b^{2}$. It is well known that this algebra  commutes with a third  screening charge
\begin{equation}
  \mathcal{W}=\oint\left(b\partial\Phi_{k}-i\beta\partial\phi_{k}\right)e^{b^{-1}(\Phi_{k}-\Phi_{k+1})}d\xi,
\end{equation}
known also as Wakimoto screening charge. It means that the theory
 \begin{equation}\label{action-dual}
   \tilde{\mathcal{A}}_{n}=\int\left(\frac{1}{8\pi}\left(\partial_{a}\Phi\cdot\partial_{a}\Phi\right)+\frac{1}{8\pi}\left(\partial_{a}\phi\cdot\partial_{a}\phi\right)+
   \tilde{\Lambda}\sum_{k=1}^{n}\left(b\partial\Phi_{k}-i\beta\partial\phi_{k}\right)\left(b\bar{\partial}\Phi_{k}-i\beta\bar{\partial}\phi_{k}\right)e^{b^{-1}(\Phi_{k}-\Phi_{k+1})}\right)\,d^{2}\xi,
\end{equation}
shares the same integrable structure in the limit $\tilde{\Lambda}\rightarrow0$ as the original one \eqref{action} in the limit $\Lambda\rightarrow0$. Actually, the theory \eqref{action-dual} makes sense only in the region $b\rightarrow\infty$. In order to regularize its UV behavior one has to add counterterms. Total renormalized action will have a form of a sigma-model 
\begin{equation}\label{action-SM}
  \tilde{\mathcal{A}}_{n}=\frac{1}{8\pi}\int G_{\mu\nu}(\boldsymbol{X}|\tilde{\Lambda},b^{2})\partial_{a}X^{\mu}\partial_{a}X^{\nu}d^{2}\xi,\quad\text{where}\quad
  \boldsymbol{X}=(\Phi_{1},\dots,\Phi_{n},\phi_{1},\dots,\phi_{n}).
\end{equation}
The precise form of the metric $G_{\mu\nu}(\boldsymbol{X}|\tilde{\Lambda},b^{2})$ might be very complicated. Moreover, it depends on a chosen regularization scheme. The only reasonable thing is to find it in the semiclassical approximation $b\rightarrow\infty$. We define
\begin{equation}
  (x_{1},\dots,x_{n},y_{1},\dots,y_{n})=b^{-1}(\Phi_{1},\dots,\Phi_{n},\phi_{1},\dots,\phi_{n}), \quad
  \tilde{\Lambda}b^{2}=e^{t},
\end{equation}
and make the following anzatz for the classical metric (here $z_{k}=x_{k}-iy_{k}$)
\begin{equation}\label{metric}
   ds^{2}=\sum_{k=1}^{n}dz_{k}d\bar{z}_{k}+\mu(t)\left(\sum_{k=1}^{n}dz_{k}\right)^{2}+
   2\sum_{k=1}^{n}dz_{k}^{2}\sum_{l=1}^{n}\lambda_{l}(t)e^{(\boldsymbol{\alpha}_{k}+\dots+\boldsymbol{\alpha}_{k+l-1},\boldsymbol{x})},
\end{equation}
where $\boldsymbol{\alpha}_{k}=\boldsymbol{\alpha}_{n+k}$ are the roots of $\mathfrak{sl}(n)$: $(\boldsymbol{\alpha}_{k},\boldsymbol{x})=x_{k}-x_{k+1}$, $x_{k+n}=x_{k}$. This anzatz is consistent with the symmetry of the problem, but of course it is a matter of a guesswork to find it.  The metric \eqref{metric} should flow with the RG time $-\infty<t<t_{0}$ according to the Ricci flow equation \cite{Friedan:1980jm}
\begin{equation}\label{Ricci-flow}
   R_{\mu\nu}+\nabla_{\mu}V_{\nu}+\nabla_{\nu}V_{\mu}=-\dot{G}_{\mu\nu},\qquad V_{\mu}=\nabla_{\mu}\Phi,
\end{equation}
where the vector field $V_{\mu}$ describes the effect of renormalization of the fields. We assume, for simplicity, that this vector field is a gradient of a scalar function: $V_{\mu}=\nabla_{\mu}\Phi$. Then there exists a  solution to \eqref{Ricci-flow} with the desired UV asymptotic 
\begin{equation}
   \mu(t)=-\frac{2}{n}\frac{e^{nt}}{e^{nt}-1},\quad
   \lambda_{k}(t)=\frac{e^{kt}}{e^{nt}-1},\qquad\Phi=\sum_{k=1}^{n}x_{k}.
\end{equation}

We note that we can perform the boost transformation
\begin{equation}
    \sum_{k=1}^{n}z_{k}\rightarrow\nu\sum_{k=1}^{n}z_{k},\quad
    \sum_{k=1}^{n}\bar{z}_{k}\rightarrow\nu^{-1}\sum_{k=1}^{n}\bar{z}_{k},
\end{equation}
and decouple the ``center of mass'' field $\sum_{k}z_{k}$ by sending  $\nu\rightarrow0$. This is equivalent to the transformation \eqref{Z-Z}, which changes the behavior of the fields in the semiclassical limit $b\rightarrow\infty$. The decoupled metric has the form
\begin{equation}\label{CPN-metric}
   ds^{2}=|d\boldsymbol{z}|^{2}+\frac{2}{e^{nt}-1}\sum_{k=1}^{n}(h_{k},d\boldsymbol{z})^{2}f_{k}(x),\qquad
      f_{k}(x)=\sum_{l=1}^{n}e^{lt}e^{(\boldsymbol{\alpha}_{k}+\dots+\boldsymbol{\alpha}_{k+l-1},\boldsymbol{x})}.
\end{equation}
This metric is in $2(n-1)$ dimensional space with complex coordinates $z_{k}=x_{k}-iy_{k}$, $k=1,\dots, n-1$, $\boldsymbol{\alpha}_{k}=\boldsymbol{\alpha}_{n+k}$ are the roots of $\mathfrak{sl}(n)$  and $\boldsymbol{h}_{k}$ are the weights of the first fundamental representation.  For example, for $n=2$ the metric \eqref{CPN-metric} reads 
\begin{equation}
  ds^{2}=dzd\bar{z}+\frac{1}{e^{2t}-1}\left((e^{t+x}+e^{2t})dz^{2}+(e^{t-x}+e^{2t})d\bar{z}^{2}\right).
\end{equation}
It can be transformed to the $T-$dual of the sausage metric \cite{Fateev:1992tk}
\begin{equation}
   ds^{2}=\frac{\kappa d\zeta^{2}}{4(1-\zeta^{2})(1-\kappa^{2}\zeta^{2})}+\frac{4(1-\kappa^{2}\zeta^{2})d\varphi^{2}}{\kappa(1-\zeta^{2})},\qquad
   \kappa=-\tanh t,
\end{equation}
by  simple change of variables
\begin{equation}
   \cosh x=\frac{1+\zeta^{2}}{1-\zeta^{2}}\qquad
   y=\frac{\varphi}{4}-i\log\left(\frac{(1-\zeta)(1+\kappa\zeta)}{(1+\zeta)(1-\kappa\zeta)}\right).
\end{equation}
It is well known that the sausage model coincides with the $\eta-$deformed $\mathbb{CP}(1)$ sigma-model.  We conjecture that our general metric \eqref{CPN-metric} coincides with the metric of the $\eta-$deformed $\mathbb{CP}(n-1)=SU(n)/SU(n-1)U(1)$ sigma-model after $T-$dualities in all isometry directions. 

The action of general $\eta$-deformed (we take $\eta=i\kappa$) $G/H$ coset sigma model has the form \cite{Delduc:2013fga}
\begin{equation}\label{Coset-action-deformed}
   \mathcal{S}=\frac{\kappa}{2}\int\textrm{Tr}\left(
   \left(\mathbf{g}\partial_{+}\mathbf{g}^{-1}\right)^{(\textrm{c})}\,\frac{1}{1-i\kappa\mathcal{R}_{\mathbf{g}}\circ\mathrm{P}_{\textrm{c}}}\,
   \left(\mathbf{g}\partial_{-}\mathbf{g}^{-1}\right)^{(\textrm{c})}\right) d^{2}x,
\end{equation}
where $\boldsymbol{g}\in G$, $\mathcal{R}_{\mathbf{g}}=\textrm{Ad}\, \mathbf{g}\circ\mathcal{R}\circ\textrm{Ad}\,\mathbf{g}^{-1}$ and $\mathrm{P}_{\textrm{c}}$ is the projection on the coset space. In our case we take $G=SU(n)$ and $H=U(n-1)=U(1)\otimes SU(n-1)$. The operator $\mathcal{R}$ acts in the Lie algebra $\mathfrak{g}=\mathfrak{c}\oplus_{\alpha>0}\mathfrak{g}_{\alpha}\oplus_{\alpha>0}\mathfrak{g}_{-\alpha}$ as 
\begin{equation}
  \mathcal{R}\Bigl|_{\mathfrak{c}}=0,\qquad
  \mathcal{R}\Bigl|_{\mathfrak{g}_{\alpha}}=i,\qquad
  \mathcal{R}\Bigl|_{\mathfrak{g}_{-\alpha}}=-i,
\end{equation}
while $\mathcal{R}_{\mathbf{g}}=A^{-1}\mathcal{R}A$, $A_{ab}=\langle t_{a}\,\mathbf{g}\,t_{b}\,\mathbf{g}^{-1}\rangle$. Consider for example $G=SU(3)$. We take the basis in the Lie algebra $\mathfrak{sl}(3)$  as
\begin{equation}
   t_{1}=\begin{pmatrix}
         0 & 1 & 0 \\
         1 & 0 & 0 \\
         0 & 0 & 0 \\
      \end{pmatrix},\quad
      t_{2}=\begin{pmatrix}
         0 & -i & 0 \\
         i & 0 & 0 \\
         0 & 0 & 0 \\
      \end{pmatrix},\quad
      t_{3}=\begin{pmatrix}
         1 & 0 & 0 \\
         0 & -1 & 0 \\
         0 & 0 & 0 \\
      \end{pmatrix},\quad
      t_{4}=\frac{1}{\sqrt{3}}\begin{pmatrix}
         1 & 0 & 0 \\
         0 & 1 & 0 \\
         0 & 0 & -2 \\
      \end{pmatrix}
\end{equation}
and
\begin{equation}
      t_{5}=\begin{pmatrix}
         0 & 0 & 1 \\
         0 & 0 & 0 \\
         1 & 0 & 0 \\
      \end{pmatrix},\quad
      t_{6}=\begin{pmatrix}
         0 & 0 & -i \\
         0 & 0 & 0 \\
         i & 0 & 0 \\
      \end{pmatrix},\quad
      t_{7}=\begin{pmatrix}
         0 & 0 & 0 \\
         0 & 0 & 1 \\
         0 & 1 & 0 \\
      \end{pmatrix},\quad
      t_{8}=\begin{pmatrix}
         0 & 0 & 0 \\
         0 & 0 & -i \\
         0 & i & 0 \\
      \end{pmatrix}.
\end{equation}
The generators $\{t_{1},t_{2},t_{3},t_{4}\}$ form a subalgebra $\mathfrak{h}=\mathfrak{su}(2)\oplus\mathfrak{u}(1)$. It is convenient to take a coset representative as
\begin{equation}
   \mathbf{g}^{-1}=e^{\frac{i(\psi-\phi)}{4}t_{3}-\frac{i(3\phi+\psi)}{4\sqrt{3}}t_{4}}e^{\frac{i\theta}{2}t_{5}}e^{i(\chi+\frac{\pi}{2})t_{7}}.
\end{equation}
For this choice $\phi$ and $\psi$ obviously correspond to isometry directions. Then, computing the metric and the $B-$field and performing $T-$dualities in $\phi$ and $\psi$ isometry directions, we find that the $B-$field vanishes while the metric takes the form 
\begin{multline}\label{CP2-sausage-metric}
   ds^{2}=\kappa\Bigl(d\chi^{2}+\frac{\sin^2\chi}{4}d\theta^2-2i\sin\theta d\theta\left(d\psi\left(\sin^2\chi-\csc^2\theta\right)+d\phi\cot\theta\csc\theta\right)-\\-
   4i\tan\chi d\chi\left(d\phi\cot^2\chi-d\psi\cos\theta\right)+
   \frac{4d\phi^2 \left((1-\kappa^2)\csc^2\theta\csc^2\chi+\kappa^2\right)}{\kappa^2}+\frac{8d\psi d\phi(\kappa^2-1)\cot\theta\csc\theta\csc^2\chi}{\kappa^2}+\\+
   \frac{d\psi^2\left(4\left(1-\kappa^2\right)\csc^2\theta\csc^2\chi+2\kappa^2\sin^2\theta\cos2\chi-2\sec^2\chi\left(\kappa^2 \cos2\theta+\kappa^2-2\right)+3\kappa^2 (\cos2\theta +3)\right)}{\kappa^2}\Bigr)
\end{multline}
This metric satisfies Ricci flow equation \eqref{Ricci-flow} with $\Phi=-\log(\sin2\chi\sin\chi\sin\theta)-8i\phi$ and $\kappa=-\tanh6 t$. It is straightforward to find how the metric \eqref{CP2-sausage-metric} is related to the metric \eqref{CPN-metric}  (here $n=3$, $q=e^{8 t}$)
\begin{equation}
\begin{gathered}
     e^{-(\boldsymbol{\alpha}_{1},\boldsymbol{x})}=q^{-1}\sin^{2}\frac{\theta}{2}\tan^{2}\chi\left(q^{\frac{3}{2}}+(1-q^{\frac{3}{2}})\cos^{2}\frac{\theta}{2}\sin^{2}\chi\right),\\
     e^{(\boldsymbol{\alpha}_{2},\boldsymbol{x})}=q^{-\frac{1}{2}}\cos^{2}\frac{\theta}{2}\tan^{2}\chi
     \left(1-(1-q^{\frac{3}{2}})\sin^{2}\frac{\theta}{2}\sin^{2}\chi\right),\\
     (\boldsymbol{\alpha}_{1},y)=8\varphi-2i\log\left(\tan\frac{\theta}{2}\left(q^{\frac{3}{2}}+(1-q^{\frac{3}{2}})\cos^{2}\frac{\theta}{2}\sin^{2}\chi\right)\right),\\
     (\boldsymbol{\alpha}_{1}+\boldsymbol{\alpha}_{2},y)=4\varphi-12\psi-2i\log\left(\sin\frac{\theta}{2}\tan\chi\left(1-(1-q^{\frac{3}{2}})\sin^{2}\frac{\theta}{2}\sin^{2}\chi\right)\right).
\end{gathered}
\end{equation}
We have checked similar statement for $\mathbb{CP}(3)$ sigma-model, thus confirming our general conjecture. The formulae in that case are too long to be presented here.

Our results verify the conjecture that $\mathfrak{gl}(n|n)$ Toda QFT approaches in the strong coupling limit the $T-$dual of the $\eta$-deformed $\mathbb{CP}(n-1)$ sigma-model. At first sight this statement looks contradictory. It is well known that $\mathbb{CP}(n-1)$ sigma-model fails to have integrability at the quantum level. Taking a look at the action \eqref{action-modified} we note that in the limit $b\rightarrow\infty$ another ``center of mass'' field $\sum_{k}\phi_{k}$ decouples, thus exactly at $b=\infty$ the theory coincides with the deformed  $\mathbb{CP}(n-1)$ sigma-model plus a free field. Taking into account loop corrections the interaction between the two parts will appear and the theory  presumably will restore the integrability. The precise mechanism of this restoration is an interesting question to be addressed.  Of course our conjecture has to be further checked by other methods in order to be fully justified. A lot such checks will be done in \cite{Fateev-cpn}.
\section*{Acknowledgments}
Some of the results of this paper were independently derived by Vladimir Fateev. The author thanks him for sharing his insights and for his kind scientific advisement. This work is  supported by the RFBR under grant 18-02-01131.

\bibliographystyle{MyStyle} 
\bibliography{MyBib}

\providecommand{\href}[2]{#2}\begingroup\raggedright\begin{thebibliography}{10}

\bibitem{Polchinski:2014mva}
J.~Polchinski, {\it {Dualities of Fields and Strings}},  {\em Stud. Hist.
  Philos. Mod. Phys.} {\bf 59} (2017) 6--20,
  [\href{http://xxx.lanl.gov/abs/1412.5704}{{\tt arXiv:1412.5704}}].

\bibitem{Fateev:2018yos}
V.~A. Fateev and A.~V. Litvinov, {\it {Integrability, Duality and Sigma
  Models}},  {\em JHEP} {\bf 11} (2018) 204,
  [\href{http://xxx.lanl.gov/abs/1804.0339}{{\tt arXiv:1804.0339}}].

\bibitem{Litvinov:2018bou}
A.~V. Litvinov and L.~A. Spodyneiko, {\it {On dual description of the deformed
  $O(N)$ sigma model}},  {\em JHEP} {\bf 11} (2018) 139,
  [\href{http://xxx.lanl.gov/abs/1804.0708}{{\tt arXiv:1804.0708}}].

\bibitem{Litvinov:2016mgi}
A.~Litvinov and L.~Spodyneiko, {\it {On W algebras commuting with a set of
  screenings}},  {\em JHEP} {\bf 11} (2016) 138,
  [\href{http://xxx.lanl.gov/abs/1609.0627}{{\tt arXiv:1609.0627}}].

\bibitem{2015arXiv151208779B}
M.~{Bershtein}, B.~{Feigin}, and G.~{Merzon}, {\it {Plane partitions with a
  ''pit'': generating functions and representation theory}},  {\em Sel. Math.
  New Ser.} {\bf 24} (2018) 21, [\href{http://xxx.lanl.gov/abs/1512.0877}{{\tt
  arXiv:1512.0877}}].

\bibitem{Prochazka:2018tlo}
T.~Proch{\'a}zka and M.~Rap{\v c}{\'a}k, {\it {$\mathcal{W}$-algebra Modules,
  Free Fields, and Gukov-Witten Defects}},
  \href{http://xxx.lanl.gov/abs/1808.0883}{{\tt arXiv:1808.0883}}.

\bibitem{Fateev-cpn}
V.~A. Fateev to appear.

\bibitem{Friedan:1980jm}
D.~H. Friedan, {\it {Nonlinear models in $2+\epsilon$ dimensions}},  {\em
  Annals Phys.} {\bf 163} (1985) 318.

\bibitem{Fateev:1992tk}
V.~A. Fateev, E.~Onofri, and A.~B. Zamolodchikov, {\it {The Sausage model
  (integrable deformations of O(3) sigma model)}},  {\em Nucl. Phys.} {\bf
  B406} (1993) 521--565.

\bibitem{Delduc:2013fga}
F.~Delduc, M.~Magro, and B.~Vicedo, {\it {On classical $q$-deformations of
  integrable sigma-models}},  {\em JHEP} {\bf 11} (2013) 192,
  [\href{http://xxx.lanl.gov/abs/1308.3581}{{\tt arXiv:1308.3581}}].

\end{thebibliography}\endgroup
\end{document}